# Core-Intermediate-Peripheral Index: Factor Analysis of Neighborhood and Shortest Paths-based Centrality Metrics

Natarajan Meghanathan, PhD
Professor of Computer Science
Jackson State University, MS
E-mail: natarajan.meghanathan@jsums.edu

**Abstract**
We perform factor analysis on the raw data of the four major neighborhood and shortest paths-based centrality metrics (Degree, Eigenvector, Betweeenness and Closeness) and propose a novel quantitative measure called the Core-Intermediate-Peripheral (CIP) Index to capture the extent with which a node could play the role of a core node (nodes at the center of a network with larger values for any centrality metric) vis-a-vis a peripheral node (nodes that exist at the periphery of a network with lower values for any centrality metric). We conduct factor analysis (varimax-based rotation of the Eigenvectors) on the transpose matrix of the raw centrality metrics dataset, with the node ids as features, under the hypothesis that there are two factors (core and peripheral) that drive the values incurred by the nodes with respect to the centrality metrics. We test our approach on a diverse suite of 12 complex real-world networks.

## 1 Introduction

The topological importance of nodes in complex networks has been analyzed in the literature from the perspectives of core-periphery structure and centrality metrics. While the core-periphery structure analysis of a network is more of a qualitative approach (and sometimes quantitative) at a mesoscopic level, centrality metrics are designed to quantify the topological importance of individual nodes in a network. The core-periphery analysis of a network is aimed at categorizing a node as either a core node or a peripheral node. The current status quo in the literature on the definitions of core nodes and peripheral nodes is that the core nodes need to be of larger degree and form a highly dense backbone to which the low degree peripheral nodes are connected to; the peripheral nodes are expected to be not connected to other peripheral nodes as well. Some of the works (e.g., [1-3]) in the literature have suggested that high degree nodes need not always be core nodes; but they still analyze the core-periphery structure and quantify the extent of coreness of a node within the realms of the above model.

Several of the existing algorithms to analyze the core-periphery structure of a network would fail to identify node 2 (in Figure 1) of our running example graph of Section 2 as a core node and on the other hand, would overrate and classify nodes 3 and 6 as core nodes. A close look at the example graph of Figure 1 would indicate that nodes 1 and 2 bring the two otherwise disconnected components of nodes ([1, 3, 5, 6, 9] and [2, 4, 7, 8, 10]) together and must be highly ranked as core nodes, whereas nodes 3 and 6 are to be treated as peripheral nodes, because their presence do not add any value to the network, other than increasing the degrees of nodes 1 and 5. We do not claim a core node has to be an articulation point [4] (whose removal would disconnect an otherwise connected graph); but, we do expect a core node to add value to the connectivity of the nodes in the network to justify the notion that the core nodes are to form the center of the network and the peripheral nodes are at the boundary of the network.

In this paper, we seek to enhance the core-periphery model by proposing a more comprehensive definition for the core nodes and peripheral nodes on the basis of the values incurred by the nodes for both neighborhood-based and shortest paths-based centrality metrics. The four commonly studied centrality metrics spanning these two categories are: Degree (DEG) [5], Eigenvector (EVC) [6], Betweenness (BWC) [7] and Closeness (CLC) [8]. While DEG (measure of the number of neighbors of a node) and EVC (measure of the degree of a node as well as the degrees of its neighbors) are neighborhood-based metrics, BWC (measure of the extent with which the shortest paths between any two nodes in the network go through the node) and CLC (measure of the lengths of the shortest paths from the node to the rest of the nodes in the network) are shortest paths-based metrics.

Our premise is: nodes that are part of the shortest paths for several node pairs and located closer to the majority (if not all) of the nodes in the network are more likely to incur reasonably larger values for the neighborhood-based centrality metrics as well. Accordingly, we propose the following definitions for a core node and a peripheral node: a core node should form the center of the network by being closer to the rest of the nodes and located on the shortest paths between several node pairs; whereas the peripheral nodes form the boundary of the network by being far away from a majority of the nodes and barely located on the shortest paths for any node pairs. Several works in the literature have focused on either qualitatively/quantitatively assessing the core-periphery structure of a network or quantitatively assessing the centrality metrics values incurred for the nodes and their correlations. To the best of our knowledge, we have not come across any work that analyzes the core-periphery structure of a network by taking into consideration the values incurred by the nodes with respect to a comprehensive set of centrality metrics spanning both the neighborhood-based and shortest paths-based categories. Our work in this paper takes the latter direction.

Our hypothesis for this research is that the values incurred by a node with respect to different centrality metrics are majorly influenced by their location: either at the core or the periphery of a network. Our hypothesis stems from the observation that the core nodes incur significantly larger values for the centrality metrics (especially, BWC, CLC and DEG), while the peripheral nodes incur significantly lower centrality values. Though there are quite a few works in the literature (e.g., [1]) that quantify the extent of coreness of a node, none of these works take into consideration the shortest paths aspect and are heavily-based on degree centrality. This forms the motivation for our research. We propose to conduct factor analysis [9] on the centrality dataset (with respect to values incurred for the DEG, EVC, BWC and CLC metrics) of the nodes in a network and seek to quantify the latent (hidden) factors (core or peripheral) that drive the values incurred by a node with respect to the different centrality metrics.

The rest of the paper is organized as follows: Section 2 explains the proposed centrality dataset-based factor analysis approach for core-periphery analysis along with a running example graph. Section 3 presents the results of running the proposed factor analysis approach on a suite of 12 diverse complex real-world networks and classifies them as heavy with respect to one or two of the three classes of nodes (core, peripheral and intermediate) that we identify from the factor analysis approach. Section 4 discusses related work in the literature and highlights the uniqueness of our work. Section 5 concludes the paper and presents plans for future work. Throughout the paper, the terms 'node' and 'vertex', 'edge' and 'link', 'network' and 'graph', 'measure' and 'metric' are used interchangeably. They mean the same.

## 2 Factor Analysis of Centrality Dataset

Factor analysis [9] is a widely used approach in machine learning to quantify the hidden factors that are behind the values incurred for the different features (columns) in a dataset (matrix) of records. For the problem in hand, the features are the nodes and we hence conduct factor analysis on the transpose matrix (that will have four rows, each pertaining to a centrality metric: DEG, EVC, BWC and CLC and *n* columns, where *n* is the number of nodes in the network) of the centrality dataset. Figure 1 presents a toy 10-node graph that we will use as a running example in this section to illustrate our proposed procedure. The transpose matrix of the centrality dataset with the 10 nodes as features is shown as well.

We first obtain the covariance matrix (see Figure 1) of the transpose matrix of the centrality dataset and determine its Eigenvalues and Eigenvectors [10]. The covariance matrix comprises of the Pearson's correlation coefficient [10] between the centrality metrics values for any two nodes. Since we hypothesize that there are only two factors behind the centrality values incurred by the nodes, we retain only the Eigenvectors (EVs; see Figure 2) corresponding to the largest and second largest Eigenvalues. The entries in these two Eigenvectors (referred to respectively as the first principal and second principal Eigenvectors) are considered as the initial loadings for the nodes (features). We build a two-dimensional coordinate system (see Figure 2) of the Eigenvectors with these node loadings as the data points.

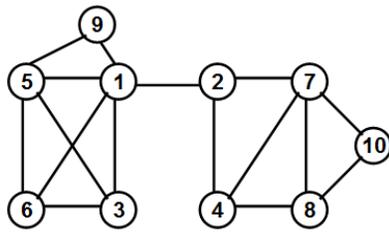

| | 1 | 2 | 3 | 4 | 5 | 6 | 7 | 8 | 9 | 10 |
|---|---|---|---|---|---|---|---|---|---|---|
| DEG | 5 | 3 | 3 | 3 | 4 | 3 | 4 | 3 | 2 | 2 |
| EVC | 0.5127 | 0.2443 | 0.3949 | 0.1564 | 0.4579 | 0.3949 | 0.1756 | 0.1208 | 0.2807 | 0.0857 |
| BWC | 21 | 20 | 0 | 3 | 1 | 0 | 9.5 | 0.5 | 0 | 0 |
| CLC | 0.0667 | 0.0667 | 0.0476 | 0.0526 | 0.0500 | 0.0476 | 0.0556 | 0.0417 | 0.0455 | 0.0400 |

Graph   Transpose Matrix of the Centrality Dataset

| | 1 | 2 | 3 | 4 | 5 | 6 | 7 | 8 | 9 | 10 |
|---|---|---|---|---|---|---|---|---|---|---|
| 1 | 1 | 0.9962 | -0.181 | 0.7471 | 0.0879 | -0.181 | 0.9806 | 0.0297 | -0.189 | -0.141 |
| 2 | 0.9962 | 1 | -0.266 | 0.6863 | 0.0003 | -0.266 | 0.9599 | -0.058 | -0.274 | -0.227 |
| 3 | -0.181 | -0.266 | 1 | 0.5161 | 0.9637 | 1 | 0.0133 | 0.9744 | 1 | 0.9959 |
| 4 | 0.7471 | 0.6863 | 0.5161 | 1 | 0.726 | 0.5161 | 0.8627 | 0.6866 | 0.5091 | 0.5526 |
| 5 | 0.0879 | 0.0003 | 0.9637 | 0.726 | 1 | 0.9637 | 0.2798 | 0.9959 | 0.9615 | 0.9713 |
| 6 | -0.181 | -0.266 | 1 | 0.5161 | 0.9637 | 1 | 0.0133 | 0.9744 | 1 | 0.9959 |
| 7 | 0.9806 | 0.9599 | 0.0133 | 0.8627 | 0.2798 | 0.0133 | 1 | 0.2246 | 0.0052 | 0.0552 |
| 8 | 0.0297 | -0.058 | 0.9744 | 0.6866 | 0.9959 | 0.9744 | 0.2246 | 1 | 0.9725 | 0.9853 |
| 9 | -0.189 | -0.274 | 1 | 0.5091 | 0.9615 | 1 | 0.0052 | 0.9725 | 1 | 0.9954 |
| 10 | -0.141 | -0.227 | 0.9959 | 0.5526 | 0.9713 | 0.9959 | 0.0552 | 0.9853 | 0.9954 | 1 |

Covariance Matrix of the Transpose Matrix

**Figure 1:** Running Example Graph; Transpose Matrix of its Centrality Dataset and its Covariance Matrix

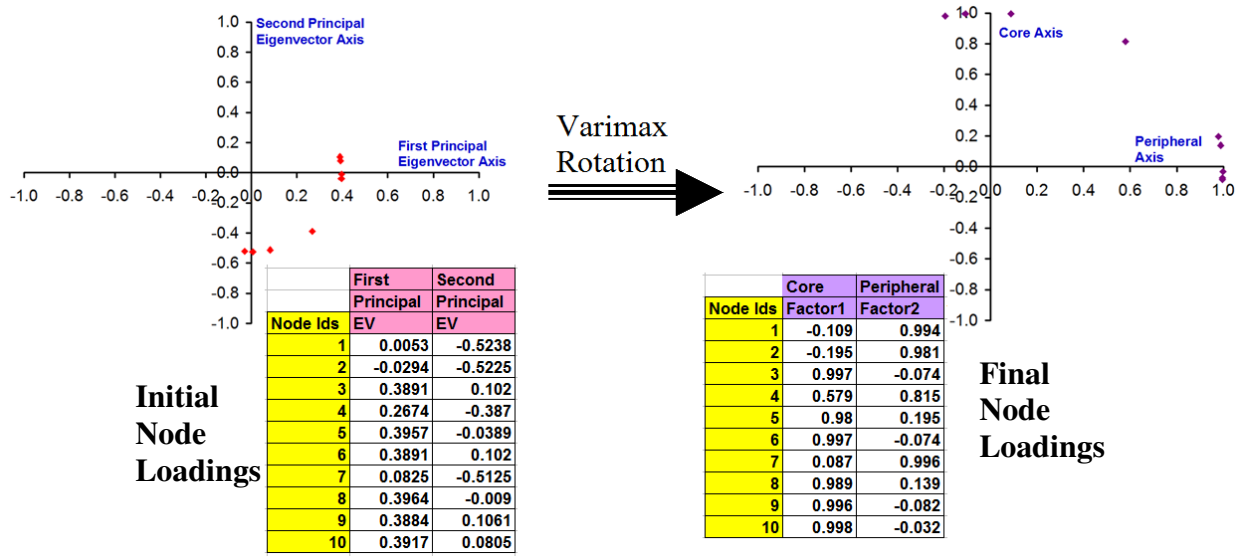

**Figure 2:** Varimax Rotation of the Eigenvector-Axes to the Factor-Axes

We seek to synchronously rotate the two Eigenvector coordinate-axes (that are orthogonal to each other) in such a way that the maximum of the data points are aligned on either of them (a procedure referred to as Varimax rotation [10] in the literature). Varimax rotation aims to maximize the communality score (sum of the squares of the loadings) for each node by going through repeated orthogonal rotations of the Eigenvector coordinate-axes. We conducted Varimax rotation using the relevant libraries available in Python (Pandas) [11]. The axes in the resulting rotated coordinate system (see Figure 2) correspond to the two factors (core and peripheral); more specifically, we treat the axis where nodes with larger values for BWC align with as the vertical Y-axis (referred to as the core-axis) and the axis where nodes with lower or zero BWC values align with as the horizontal X-axis (referred to as the peripheral-axis). The coordinates (final loadings; see Figure 2) of the nodes in such a rotated coordinate system are expected to be either close to (1, 0): if the node is a peripheral node or (0, 1): if the node is a core node.

**Core-Intermediate-Peripheral (CIP) Index:** When we conducted the above-described procedure for factor analysis on the centrality datasets of several complex real-world networks, we observed the density and pattern of the distribution of the final loadings (coordinates) of the nodes in the peripheral axis-core axis coordinate system to depend on whether the network is a random network [12] or a scale-free network [13] or in between these two extreme categories. This motivated us to determine the tan (tangent) angle (referred to as the Core Intermediate Peripheral: CIP Index) of the line joining the origin (0, 0) and the coordinates for a node in the peripheral axis-core axis coordinate system. The angle is measured anti-clockwise from the peripheral-axis for those coordinates lying in the first and second quadrants, and measure clockwise from the peripheral-axis for those coordinates lying in the fourth quadrant.

With the core-axis being the vertical Y-axis (and the peripheral-axis being the horizontal X-axis), the CIP index value for a node could be construed as a quantitative estimate of the extent with which a node could serve as a core node in the network. If the CIP index value for a node is closer to 90 degrees, then the node could be considered a core node and if the CIP index value for a node is closer to 0 degrees, the node could be considered a peripheral node. Figure 3 illustrates the CIP angle measurements for nodes 2 (a core node), 4 (an intermediate node) and 9 (a peripheral node) in the example graph.

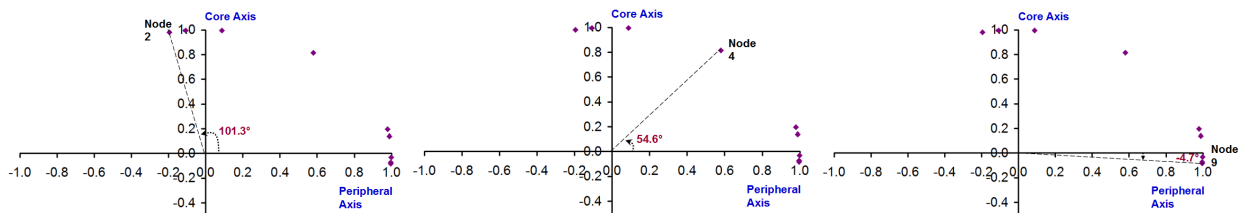

**Figure 3:** Sample Core-Intermediate-Peripheral (CIP) Index Angle Measurements for Nodes in the Example Graph

**3-element CIP Bins-Fraction Tuple:** We also propose a binning approach (with interval of 10 degrees: < 0, 0..10, 10...20, etc.., 80...90 and > 90) to group the nodes based on their CIP values. We treat the bins for CIP values falling in either of these two ranges [80,..., 90) and [90, ....) as those corresponding to the core nodes and the bins for CIP values falling in either of these two ranges [..., 0) and [0, ..., 10) to correspond to the peripheral nodes. We categorize the rest of the CIP bins to belong to nodes that are neither core nor peripheral (we refer to such nodes as intermediate nodes). We observe the centrality values for such intermediate nodes to be neither too high nor too low (i.e., not negligible, but not appreciably high as well). For a given network (of $n$ nodes), we determine a 3-element CIP bins-fraction tuple that would comprise the fractions of the nodes in the core class, peripheral class and intermediate class. If the value for any of these three fractions is greater than or equal to 0.5, we categorize the network as heavy with respect to that class (i.e., core-heavy or peripheral-heavy or intermediate-heavy). If all the three fractions in the 3-element CIP bins-fraction tuple for a network are less than 0.5, we categorize the network as heavy with respect to the classes with the top two fraction values. We observe the random networks to be typically intermediate-heavy; scale-free networks with a lower variation of node degree to be core-heavy on its own or core/intermediate-heavy or core/peripheral-heavy and scale-free networks with a larger variation of node degree to be peripheral-heavy.

Figure 4 presents the CIP index values for all the nodes in the network as well as ranks them on the basis of the extent of coreness. To the best of our knowledge, this is the first such work to be able to individually rank the nodes with respect to the extent of coreness in the form of a real-valued quantitative measure taking into consideration both the neighborhood-based and shortest paths-based centrality metrics (unlike the k-core measure [14] and its variants [1] that can take only integer values and are also insensitive to the shortest paths-based centrality metrics). Figure 4 also presents a count of the number of nodes falling in each of the CIP index range bins: we observe 3 of the 10 nodes to be core, 2 nodes to be intermediate and 5 nodes to be peripheral. The 3-element CIP bins-fraction tuple for the example graph is thus [3/10, 2/10, 5/10] = [0.3, 0.2, 0.5]. Since the fraction of peripheral nodes satisfies the criteria of being 0.5 or above, we conclude the example graph to be peripheral-heavy. In addition, Figure 4 also visually illustrates the core-peripheral structure of the graph in a typical layout as well as using the Yifan proportional layout algorithm [15] (in both the layouts, the node color corresponds to the class of the node). In the Yifan Hu proportional layout, the node size is proportional to their CIP index values as well.

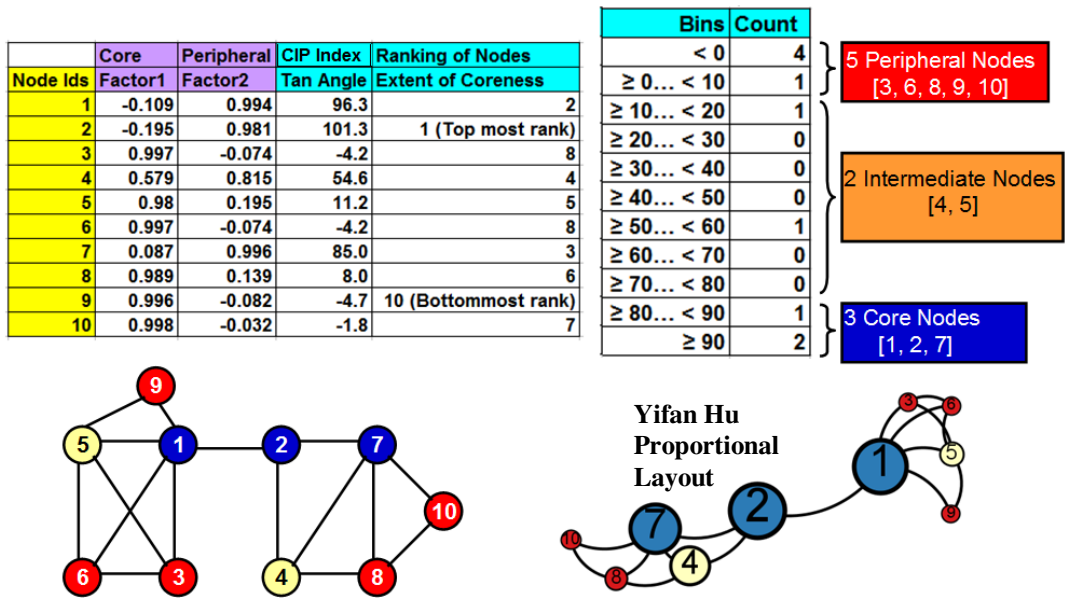

**Figure 4:** Ranking, Classification and Visualization of the Nodes as Core, Intermediate and Peripheral Nodes based on their CIP Index Values

**Comparison with *k*-Core:** From the perspective of the procedure (see Section 4) to determine the *k*-core measure, all the four nodes 1, 3, 5 and 6 in the example graph of Section 2 would form a 3-core graph, the highly-dense sub graph possible in the graph. But, per the CIP index measure proposed in this paper: node 1 is the only core node among these four nodes, whereas nodes 3 and 6 are peripheral nodes. We see the CIP index-based classification of these four nodes to more appropriately fit our definition of core nodes and peripheral nodes stated in Section 1. None of the node pairs in the network go through nodes 3 or 6 for shortest paths as well as these two nodes are relatively (compared to node 1) farther away from the rest of the nodes in the network. On the other hand, we see the classification of nodes 1, 2 and 7 as core nodes (forming the center of the network) is well-justified by the visualization obtained using the Yifan Hu proportional layout algorithm.

## 3 Evaluation with Real-World Networks

In this section, we present the results obtained by running the proposed factor analysis procedure of Section 2 on a suite of 12 diverse real-world networks, ranging from pure random networks to extremely scale-free networks. The diversity among these networks is captured in the form of the spectral radius ratio for node degree (represented as $\lambda_{sp}$) [16], a metric that seamlessly captures the variation in node degree independent of the number of nodes and edges in the network. The $\lambda_{sp}$ value [16] for a network is the ratio of the largest Eigenvalue of the 0-1 adjacency matrix of the network and the average node degree. Table 1 presents the name, number of nodes and edges and the $\lambda_{sp}$ value of the networks as well as the results (the 3-element bins-fraction tuple and the network classification) for the 12 real-world networks. The networks in Table 1 are ordered in the increasing order of their $\lambda_{sp}$ values. Most of these real-world networks have been used as benchmark networks in several studies related to complex network analysis.

**Table 1:** Real-World Networks used for Centrality-based Factor Analysis and their 3-Element Bins CIP Fractions Tuple-based Classification

| Net # | Network Name | # Nodes | # Edges | $\lambda_{sp}$ | 3-Element Bins Fraction Tuple [C, I, P] Fractions | Network Classification |
|---|---|---|---|---|---|---|
| Net 1 | US Football Net | 115 | 613 | 1.01 | [0.00, 1.00, 0.00] | Intermediate-heavy |
| Net 2 | Taro Exchange Net | 22 | 39 | 1.06 | [0.36, 0.59, 0.05] | Intermediate-heavy |
| Net 3 | Flying Teams Cadets Net | 48 | 170 | 1.21 | [0.50, 0.35, 0.15] | Core-heavy |
| Net 4 | Dolphin Net | 62 | 159 | 1.40 | [0.68, 0.16, 0.16] | Core-heavy |
| Net 5 | Band Jazz Net | 198 | 2742 | 1.44 | [0.33, 0.46, 0.21] | Intermediate/Core-heavy |
| Net 6 | Karate Net | 34 | 78 | 1.47 | [0.35, 0.21, 0.44] | Peripheral/Core-heavy |
| Net 7 | Adjacency Noun Net | 112 | 425 | 1.73 | [0.62, 0.26, 0.12] | Core-heavy |
| Net 8 | Les Miserables Net | 77 | 254 | 1.82 | [0.28, 0.14, 0.58] | Peripheral-heavy |
| Net 9 | Copper Field Net | 87 | 406 | 1.83 | [0.11, 0.39, 0.50] | Peripheral-heavy |
| Net 10 | Anna Karenina Net | 138 | 493 | 2.48 | [0.22, 0.10, 0.68] | Peripheral-heavy |
| Net 11 | US Airports 1997 Net | 332 | 2126 | 3.22 | [0.27, 0.18, 0.55] | Peripheral-heavy |
| Net 12 | EU Air Transport Net | 405 | 1981 | 3.81 | [0.24, 0.21, 0.55] | Peripheral-heavy |

We use the magnitude of the entries in the 3-element bins-fraction tuple as the basis to classify a network: If the fraction of nodes for one of the three classes is greater than or equal to 0.50, we classify the network as heavy with respect to the particular class only (i.e., either core-heavy or intermediate-heavy or peripheral-heavy). If all the three fractions of nodes are less than 0.50, we classify the network as heavy with respect to the top two classes (referred to in the decreasing order of the fraction values). For example, the well-studied Karate Network (Net 6) incurs a 3-element bins fraction tuple of [0.35, 0.21, 0.44] and accordingly the network is classified as peripheral/core-heavy: implying that the largest fraction of nodes in the network are peripheral nodes and the next largest fraction of nodes are the core nodes.

We observe the random networks (US Football Network and the Taro Exchange Network) to be intermediate-heavy. The rest of the real-world networks are scale-free networks with different levels of variation in node degree. We observe the scale-free networks with relatively lower $\lambda_{sp}$ values to be predominantly core-heavy (mostly on their own or sometimes in association with either intermediate nodes or peripheral nodes). Among the five real-world networks, with $\lambda_{sp}$ values ranging from 1.21 to 1.73, we observe three of them to be core-heavy, one network to be intermediate/core-heavy and the other network to be peripheral/core-heavy. On the other hand, all the five scale-free real-world networks with a relatively larger variation in node degree (i.e., with $\lambda_{sp}$ values ≥ 1.82) are observed to be peripheral-heavy. These are significant observations that have not been hitherto reported in the literature. In Figure 5, we notice the intermediate nodes to be typically nodes that are adjacent to the core nodes and/or the peripheral nodes; they may not be at the center of the network, but may not be at the boundary of the network as well. It is very important to recognize such nodes and categorize them to a separate class rather than strictly following the core-peripheral two-layer model.

Figure 5 presents a visualization of each of the 12 real-world networks (identified with the Net #s used in Table 1), in the increasing order of their $\lambda_{sp}$ values. All the real-world networks are displayed per the Yifan Hu proportional layout algorithm run in Gephi [17]. We also follow the same coloring convention for the core, intermediate and peripheral nodes as is shown in Figure 4 (i.e., the core nodes are colored in blue; the peripheral nodes are colored in red, and the intermediate nodes are colored in Whitish-Yellow). We observe the real-world networks to showcase the trend explained above (i.e., random networks are intermediate-heavy; scale-free networks with lower $\lambda_{sp}$ values are core-heavy on their own or in association with an another class; scale-free networks with larger $\lambda_{sp}$ values are peripheral-heavy). The two airport networks (US and EU) are peripheral-heavy and this is understandable with the presence of several stub airports compared to the hub airports (even though the number of hub airports is not negligible, vindicating the scale-free degree distribution of these networks). We also observe that for four of the five peripheral-heavy networks (including the two airport networks), the next largest fraction of nodes are the core nodes and not the intermediate nodes.

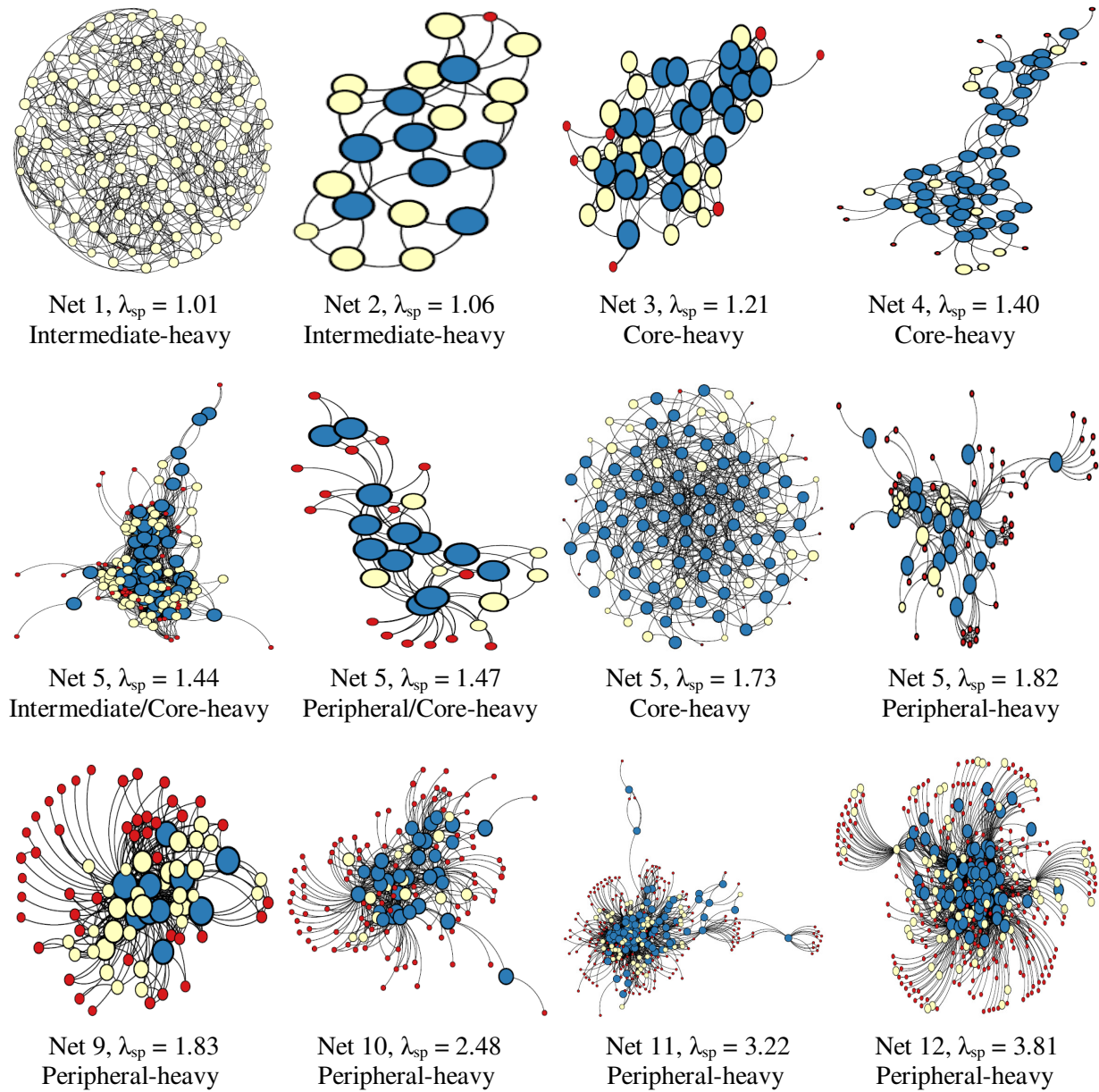

**Figure 5:** Yifan Hu Proportional Layout Algorithm-based Visualization of the Real-World Networks [Node Size and Color set per the CIP Index Values and the Node Classification]

Note that the Yifan Hu proportional layout algorithm was run on the real-world networks without any inputs regarding the CIP index values for the nodes. Only the node size and color were set per the CIP index value and the network classification respectively. The presence of the blue-colored nodes at the core/center of the network and the red-colored nodes at the periphery of the network per the Yifan Hu proportional layout algorithm is an implicit validation of our approach.

## 4 Related Work

Borgatti and Everett [18] formulated a hub-and-spoke structure for the core-peripheral nodes in a network: the core nodes are mainly adjacent to other core nodes and in some instances adjacent to peripheral nodes; whereas, peripheral nodes do not connect with other peripheral nodes. The above model formulation fits mainly for scale-free networks, but not for random networks. On the other hand, the *k*-core measure [14, 19] divides the network to a layered hierarchy and is by far the most commonly used centrality metric in the literature to capture the coreness of a node. The *k*-core of a graph is determined by removing all nodes whose degrees are less than *k* and this procedure is repeatedly applied on the graphs resulting from the node removals; the procedure is stopped when all the remaining nodes in the graph (forming a connected sub graph) have degree greater than or equal to *k*. The *k*-core value for all the nodes in the sub graph is *k*. The above procedure is used repeatedly (with different values of *k*) to find the *k*-core values for all the nodes in the network. The *k*-core measure is strongly-correlated to the EVC (eigenvector centrality) of the nodes; but, it does not take into account the participation of the nodes in the shortest paths between any two nodes as well as the proximity of the nodes to the rest of the nodes in the network. The procedure for determining the *k*-core measure is thence vulnerable (like we observed in the case of the running example graph of Section 2) to find a highly dense sub graph (for a larger *k* value) that may be farther away from the rest of the nodes in the network. Besides, the *k*-core value for a node will be just an integer and several nodes could incur the same value of *k*. Hence, the *k*-core procedure could lead to ambiguity in the ranking of the nodes (two or more nodes might incur the same *k*-core value). In 2010, a centrality metric referred to as *coreness* [1] of a node was proposed and formulated as the sum of the *k*-core measures of the neighbors of the node. The above-described *k*-core procedure is ideally suited for undirected and unit-weight graphs. Various modified versions of this procedure have been proposed in the literature, especially with respect to weighted graphs. Garas et al [20] define the weighted degree of a node as the sum of its degree and the weights of the edges incident on the node. Proposals (e.g., [21-22]) that weigh these two terms (node degree and the weights of the incident edges) are also available in the literature.

Our proposal to introduce the notion of intermediate nodes also fits within the k-core idea. The contribution of non-core nodes (the intermediate nodes and peripheral nodes in our jargon) has been recognized in [23] for the effective spread of social media protests from the core nodes (forming the epicenter) to the rest of the nodes in the network; though the non-core nodes may not have as many neighbors as the core nodes, the sheer number of such non-core nodes (especially for scale-free networks) was stated to be critical for information diffusion. Likewise, in the study of human brain dynamics, the authors [24] note that the separation between the stiff temporal core region (composed primarily of sensorimotor and visual regions) and the flexible temporal peripheral region (composed primarily of multimodal association regions) is important to assess/predict the individual differences in learning success. In [1], the authors state that the most influential spreaders in a social network may not be the most central nodes with high BWC; thus, acknowledging the contribution of the non-core nodes for information diffusion and epidemic spread. In [2-3], the authors advocated a core-periphery structure for networks wherein there are multiple cores and multiple peripheries, rather than a single hub-and-spoke model. This in turn implies that not all cores and peripheries are of the same size and density, indicating a need for a third class of nodes (the intermediate nodes).

We now review some of the methodologies (other than those based on the *k*-core measure) available in the literature to determine the core-periphery structure. In [25], the authors differentiate core-periphery structure from assortative mixing or dissortative mixing [26] by stating that the probability of connection between two core nodes is greater than the probability of connection between a core node and a peripheral node, which is in turn greater than the probability of connection between two peripheral nodes. The above rule formed the basis of their stochastic block model (involving the method of maximum likelihood) to determine the core-periphery structure. In [27], the authors use a random walker approach to build the core-periphery profile of the nodes in a network and use the measure of persistence probability (the fraction of time a walker would spend on a core node across the entire walk) as a quantitative measure of the coreness of the node. Note that the CIP index value proposed in our work is a deterministic quantity and captures the exact level of coreness for a node. In [28], the authors propose a motif-based approach that involves spectral analysis of the motif-adjacency matrix to identify the core nodes and periphery nodes of a network such that the degree of the core nodes (and peripheral nodes) are greater (lower) than the average degree; still, this approach is only degree-based and does not take into account the shortest paths-based centrality metrics. The minimum residual method [29] seeks to assign both the end vertices of an edge $(i, j)$ to either core or peripheral sets such that the square of the number of residuals is minimized. A residual for a pair of vertices $(i, j)$ is a 1 if there is an edge between $i$ and $j$, but $i$ and $j$ are not in the same grouping (core or periphery) or a -1 if there is no edge between $i$ and $j$, but $i$ and $j$ are in the same grouping of vertices.

## 5  Conclusions and Future Work

Our contributions in this paper are the following: (1) We propose a quantitative measure (referred to as Core-Intermediate-Peripheral: CIP Index) to capture the extent with which a node could serve as core node in a network. Unless two nodes incur the same values for all the four centrality metrics (DEG, EVC, BWC, CLC) based on which the CIP index values for the nodes are determined, the CIP index values (a real-valued measure) for any two nodes are expected to be different and could be used to unambiguously rank the nodes in the network with respect to the extent the nodes could serve as core nodes. (2) Rather than two classes (core or peripheral), we propose three classes of nodes (core, intermediate and peripheral) and the CIP index value incurred for a node could be used to assign the class to which a node belongs to. The 3-element CIP bins-fraction tuple determined for a network could be used to decide whether the network is heavy with respect to a particular class or two of the three classes. (3) Upon evaluation of the centrality metrics datasets of a suite of 12 diverse real-world networks, we observe random networks to be intermediate-heavy; scale-free networks with lower variation in node degree to be core-heavy on their own or in association with an another class (intermediate or peripheral) and scale-free networks with relatively larger variation in node degree to be peripheral-heavy. As part of future work, we plan to conduct factor analysis on centrality metrics datasets of real-world networks with the above four centrality metrics as features and determine the number of latent factors behind the centrality metric values incurred by nodes in random networks and scale-free networks with lower and higher variations in node degree.

**Acknowledgments**
The work leading to this paper was partly funded through the U.S. National Science Foundation (NSF) grant OAC-1835439. The views and conclusions contained in this paper are those of the authors and do not represent the official policies, either expressed or implied, of the funding agency.